# Heat transfer at the van der Waals interface between graphene and NbSe₂

**Yohta Sata[1], Rai Moriya[1,*], Naoto Yabuki[1], Satoru Masubuchi[1], and Tomoki Machida[1,*]**

[1]Institute of Industrial Science, University of Tokyo, 4-6-1 Komaba, Meguro, Tokyo 153-8505, Japan

E-mail: moriyar@iis.u-tokyo.ac.jp, tmachida@iis.u-tokyo.ac.jp

**Abstract**. Graphene has been widely used to construct low-resistance van der Waals (vdW) contacts to other two-dimensional (2D) materials. However, a rise of graphene's electron temperature under a current flow has not been seriously considered in many applications. Owing to its small electronic heat capacity and electron-phonon coupling, graphene's electron temperature can be increased easily by the application of current. The heat generated within the graphene is transferred to the contacted 2D materials through the vdW interface and potentially influences their properties. Here, we compare the superconducting critical currents of an NbSe₂ flake for two different methods of current application: with a Au/Ti electrode fabricated by thermal evaporation and with a graphene electrode contacted to the NbSe₂ flake through a vdW interface. The influence of the heat transfer from the graphene to NbSe₂ is detected through the change of the superconductivity of NbSe₂. We found that the critical current of NbSe₂ significantly reduces when the current is applied with the graphene electrode compared to that from the conventional Au/Ti electrode. Further, since the electron heating in graphene exhibits ambipolar back-gate modulation, we demonstrate the electric field modulation of the critical current in NbSe₂ when the current is applied with graphene electrode. These results are attributed to the significant heat transfer from the graphene electrode to NbSe₂ through vdW interface.

## 1. Introduction

Graphene has been widely studied as a high-quality electrode material for van der Waals (vdW) heterostructures. Exfoliated graphene exhibits an inert surface owing to its non-bonding nature and can be transferred easily onto other two-dimensional (2D) materials to construct highly transparent vdW contact. Various high-quality 2D material heterostructures have been achieved with the help of graphene vdW contact; for example, high mobility $MoS_2$ transistors [1,2], air-stable black phosphorus [3,4], light-emitting diodes based on vdW heterostructures [5], and vertical field-effect transistors based on graphene/transition metal dichalcogenide (TMD) vdW heterostructures [6,7]. More recently, owing to its small electronic heat capacity and small electron–phonon coupling, graphene has received considerable attention for studying hot carrier dynamics such as photodetection [8], thermal emission [9], bolometric effects [10], multiple hot carrier generation [11], and heat transfer at vdW interface [12,13]. In these experiments, heat dissipation pathways of hot electrons within graphene are important because they are used to determine the performance of the above-mentioned devices. In particular, recent publications revealed that out-of-plane electronic heat transfer in the vdW interface is a dominant pathway of heat dissipation in vdW heterostructures [12,14]. Such an out-of-plane electron heat transfer might overcome intrinsically weak phonon heat transfer at the vdW interface and could be useful to control the local heat flow and thermoelectric property in vdW heterostructures between graphene and other 2D materials [15,16]. Thus far, vertical heat transport in vdW heterostructure was detected using optical method [12-14]. However, another detection scheme such as electrical detection is required for device applications. Since superconductivity is very sensitive to temperature, a change in the superconductivity can be used as a tool to detect heat transfer between a superconductor and an adjacent material [17-19]. In this study, we investigate heat transfer in a vdW interface by using a graphene/superconductor junction. $NbSe_2$ is selected as a superconductor material because this material demonstrated the ability to construct high-quality vdW heterostructures with other 2D materials [4,20-22].

## 2. Method

An optical micrograph of the fabricated mono-layer graphene/23-nm-thick NbSe$_2$ device is shown in Fig. 1(a). NbSe$_2$ devices with two different types of contacts are fabricated: a conventional thermally evaporated contact of Au/Ti metal stack electrode and the vdW contact of the graphene electrode. To fabricate this device, we adopted a mechanical exfoliation and dry transfer technique. First, mono- to few-layer graphene was fabricated onto the 300-nm-thick SiO$_2$/highly-doped-Si substrate. Separately, a flake of NbSe$_2$ with the thickness range 23–51 nm were fabricated onto a polymer sheet (Gel-Pak, PF-X4). By using a dry transfer method, the NbSe$_2$ flake was transferred on graphene to construct the graphene/NbSe$_2$ vdW junction between freshly cleaved surfaces [21,23]. Subsequently, electron beam (EB) lithography and EB evaporation was used to form Au (30 nm)/Ti (50 nm) electrodes numbered from 1 to 6 in Fig. 1(a) on both graphene and NbSe$_2$ flakes. The Au/Ti electrode number 5 was broken during metal lift-off process, thus did not use for the measurement. The fabrication was carried out without introducing any heat treatment to avoid degrading the NbSe$_2$ flake [21,24]; using same fabrication procedure, high quality NbSe$_2$-based vdW junction without any oxidation at interface has been previously demonstrated [21]. To measure the current–voltage ($I$–$V$) characteristics of the NbSe$_2$, the current $I$ was applied between the contacts, and then, the voltage difference was measured using a voltmeter with four-terminal geometry when there was a voltage drop within the NbSe$_2$. For the $dV/dI$ measurement, ac-current $I_{AC}$ = 10 nA with a frequency of 18 Hz was applied, and then ac-voltage was measured using a lock-in amplifier. The back-gate voltage ($V_{BG}$) was applied to the highly doped-Si substrate to change the carrier density of graphene. The transport properties were measured using the variable temperature cryostat.

## 3. Results

First, we compared current–voltage ($I$–$V$) characteristics between two different geometries as illustrated in Figs. 1(b) and 1(c); here we illustrated the direction of current flow in NbSe$_2$ and the position of electrical contacts to detect the voltage drop within NbSe$_2$ layer. These geometries are current application with the graphene electrode (Fig. 1(b)) and with the metal (Au/Ti) electrode (Fig. 1(c)). For current application with the graphene electrode (Fig. 1(b)), current is applied between the graphene and Au/Ti contacts (between terminals 3 and 1 in Fig. 1(a)). The electron temperature of the graphene increases significantly during the current application due to its small heat capacity and electron-phonon coupling [8,10]. The heat flow illustrated in Fig. 1(d) includes heat conduction through electron diffusion at the graphene/NbSe$_2$ vdW interface $G_e^{\text{vdW}}$, heat conduction from electron to phonon in graphene $G_{\text{ep}}^{\text{Gr}}$, heat conduction through phonons at the graphene/NbSe$_2$ vdW interface $G_p^{\text{vdW}}$, heat conduction between electron to phonon in NbSe$_2$ $G_{\text{ep}}^{\text{NS}}$, and heat conduction from graphene or NbSe$_2$ to heat bath (SiO$_2$/Si substrate) via phonon. Here, we assume that electron-phonon coupling in NbSe$_2$ is strong enough such that they are quickly thermalized with each other. Part of the heat generated in graphene is transferred to the NbSe$_2$ layer through the vdW interface between graphene and NbSe$_2$. As can be seen from the figure, there are two possible cooling paths for the electron temperature of the graphene: electron heat transfer $G_e^{\text{vdW}}$ and electron-phonon coupling $G_{\text{ep}}^{\text{Gr}}$. Through one or both of these cooling paths, the heat transferred to NbSe$_2$ subsequently influences the superconducting property of NbSe$_2$ and detected through the resistance change of NbSe$_2$. For current application with the metal electrode (Fig. 1(c)), the current is applied between two Au/Ti contacts (between terminals 3 and 6 in Fig. 1(a)) and the voltage difference between terminals 2 and 1 is measured. Here, temperature rise in Au/Ti during current application is negligibly small due to the large heat capacity of the Au and Ti metals. In this case, the dominant contribution of the applied current to the superconductivity of the NbSe$_2$ is only the conventional Oersted field effect, such that magnetic field generated by the current breaks the superconductivity when its value exceeds the critical field. By comparing the two geometries, the influence of the heat transfer at vdW interface could be determined. Note that the contact resistances between the Au/Ti and the NbSe$_2$ in the device shown in Fig. 1(a) are ranged between 40–90 Ω measured by applying voltage to each of the Au/Ti contacts while other contacts are connected to electrical ground. From the quantum Hall effect measurement, the contact resistance of NbSe$_2$/graphene contacts is determined smaller than ~50 Ω (see Appendix A).

The $I$–$V$ curves measured in different geometries obtained from mono-layer graphene/ NbSe$_2$ device are shown in Fig. 2(a). In the figure, the measurement results at 2.0 K and 7.8 K are plotted as solid and dashed lines, respectively. Here, 2.0 K and 7.8 K are below and above the critical temperature $T_c$ ~ 6.8 K of NbSe$_2$, respectively. We set $V_{BG}$ = −50 V to ensure highly hole-doped graphene. In both geometries, when increasing current at 2.0 K, measured voltage (or resistance of the NbSe$_2$) deviates from zero to finite value above their critical current $I_c$; suggesting breakdown of the superconductivity. With further increase of the current, NbSe$_2$ turns into the normal-metal state and the $I$–$V$ curve follows a linear relationship. We note that the $I$–$V$ curves for the normal states are identical to that measured at 7.8 K, which is a higher temperature than the critical temperature $T_c$ of NbSe$_2$. Noticeably, we obtained significantly different critical current ($I_c$) values between the two geometries such that $I_c$ ~ 400 μA for current application with the graphene electrode and $I_c$ ~ 0.95 mA for current application with the metal electrode. Further, the $I_c$ values measured with different voltage probe configurations are compared and the results are presented in Figs. 2(b) and 2(c) for current application with graphene and metal contacts, respectively ($I$-$V$ curves measured using different voltage probe configurations are presented in the Appendix B). The $I_c$ values are nearly the same under the same current application geometry irrespective to the voltage probe configurations. We identified that $I_c$ is smaller for current application with the graphene electrode than that with the metal electrode. These results suggest that the graphene vdW contact strongly influences the superconducting property of the NbSe$_2$. The critical current density is calculated as $8 \times 10^5$ A/cm$^2$ for the current application with the metal electrode. The critical current density measured with this contact geometry is in good agreement with that typically observed in NbSe$_2$ due to the Oersted field effect of the current [25].

The resistance of graphene electrode can be significantly tuned by $V_{BG}$. The $I$–$V$ curves measured at 2.0 K with different $V_{BG}$ values are compared in both contact geometries and the results are shown in Fig. 3(a) and 3(b). In Fig. 3(a), $I$–$V$ curves measured at different $V_{BG}$ values are plotted for current application with the graphene electrode. We observed a significant change of $I$–$V$ curves with varying $V_{BG}$. $I_c$ is larger at $V_{BG} \pm 50$ V and it exhibits a significant drop when $V_{BG} \approx 0$ V. In contrast, we did not observe a noticeable difference between the $I$–$V$ curves measured at different $V_{BG}$ values for the current application with the metal electrode as shown in Fig. 3(b). The change of $I_c$ with $V_{BG}$ for current application with the graphene electrode can be more clearly seen in Fig. 3(c), where $I_c$ vs. $V_{BG}$ is plotted at 2.0 K. We demonstrated the lowest $I_c$ of ~100 μA at $V_{BG} = 0$ V, which is nearly an order of magnitude reduction for $I_c$ compared to the current application with the metal electrode (Fig. 3(c)). For comparison, the $V_{BG}$ dependence of the two-terminal resistance of graphene is determined by applying a current between terminals 3 and 1, and measuring the voltage between terminals 2 and 1 and the result is shown in Fig. 3(d) (measurement geometry is also depicted in the inset of the figure). It shows the ambipolar modulation of the graphene's resistance with charge neutrality point at $V_{BG} \sim 0$ V. We found that Figs. 3(c) and 3(d) indicate good agreement with each other. This suggests that the modulation of $I_c$ with $V_{BG}$ for current application with the graphene electrode is originated from the change of the resistance of graphene electrode. These results contrast with the relation between $I_c$ and $V_{BG}$ for the case of current application with the metal contact as shown in Fig. 3(e). The result shows an $I_c \sim \pm 0.95$ mA irrespective of $V_{BG}$; therefore, we think there is no apparent electric field effect in this case. This seems to be in good coincidence with the fact that the electric field effect on the Au/Ti is negligibly small in the $V_{BG}$ range we applied.

From these results, we infer that the resistance of the graphene contact plays a significant role for the reduction and the $V_{BG}$ modulation of the $I_c$. Different resistances of the graphene contact result in different amounts of Joule heating under the application of current. Therefore, we estimated the total power $P$ injected into the graphene with the measurement geometry shown in Fig. 4(a) (note that this is same geometry as Fig. 3(d)). In this geometry, contact resistances of Au/Ti/graphene, NbSe$_2$/graphene contacts, and resistance of graphene are measured in series. However, since these contact resistances are much smaller than the resistance of graphene, most of the injected power is dissipated in the graphene region (see Appendix A). The $I$-$V$ curves measured in different $V_{BG}$ values are shown in Fig. 4(b). The curves are offset for clarity with the dashed lines denoting the $V = 0$ V levels for each curve. From these $I$-$V$ curves, current $I$ dependence of the injected power $P = I \times V$ for different $V_{BG}$ values are determined and plotted in Fig. 4(c); $P$ quadratically increases with increasing $I$. When $V_{BG}$ is changed from $\pm 50$ V to 0 V, $P$ increases more rapidly with $I$. On each curve, we marked the point for $P = 50$ μW with solid circles for clarity. Since the resistance of graphene increases as it approaches its charge neutrality point, the current value $I$ for the same injected power $P$ becomes lower towards $V_{BG} = 0$ V and higher towards $V_{BG} = \pm 50$ V. The current values $I$ for the same injected powers of $P = 25$, 50, and 75 μW are plotted for different $V_{BG}$ in Fig. 4(d). For comparison, the change of $I_c$ of the NbSe$_2$ layer with respect to $V_{BG}$ for current application with the graphene electrode at 2.0 K (data presented in Fig. 3(c)) was plotted together. The $I_c$ vs. $V_{BG}$ data shows good agreement with the current value for the constant power of $P = 50$ μW. Therefore, it is suggested that the breakdown of superconductivity in NbSe$_2$ with current application with the graphene electrode is determined by the injected power. This is in contrast from conventional current-induced breakdown of superconductivity, where breakdown is determined by the critical current density. These results suggest that the breakdown of superconductivity in NbSe$_2$ with current application with the graphene electrode is due to the heating of the graphene electrode.

By changing the measurement temperature, the phonon temperature of both graphene and NbSe$_2$ is controlled. Thus, the temperature dependence of $I$-$V$ curve provides us with information on the contribution of phonon temperature. In Figs. 5(a)-5(c), the temperature dependence of $I$-$V$ curves measured with $V_{BG}$ values of $-50$, $-15$, and 0 V are shown. Irrespective of the $V_{BG}$ value, $I_c$ decreases as the measurement temperature approaches the critical temperature $T_c$ of NbSe$_2$ ($\sim 6.8$

K). Similarly to the previous section, we extracted power injected in the graphene, $P$, as a function of injected current $I$ at each $V_{BG}$ and temperature. Then, the $I$–$V$ curves in Figs. 5(a)-5(c) are replotted as $P$–$V$ as shown in Figs. 5(d)-5(f). In the figure, $+P$ denotes the injected power for current flow from NbSe$_2$ to graphene and -$P$ denotes the injected power for current from graphene to NbSe$_2$. The overall changes of the $P$–$V$ curve in Figs. 5(d)-5(f) are nearly identical irrespective of the $V_{BG}$ values. This suggests that the current-induced breakdown of the NbSe$_2$ in this case is solely determined by the electron heating of graphene within the measurement temperature range for both doped graphene ($V_{BG}$ is away from 0 V) and charge-neutral graphene ($V_{BG} = 0$ V) case. From Figs. 5(d)-5(f), critical power $P_c$ is defined such that above this power value, the measured voltage (or resistance of the NbSe$_2$) deviates from zero to a finite value (e.g. $P_c$ values for the case of positive $P$ is indicated by the arrows in the figure). Then, the $P_c$ as a function of temperature $T$ is plotted for different $V_{BG}$ values of $-50$, $-15$, and 0 V in Fig. 5(g). For all the $V_{BG}$ values, the $P_c$ decreases with increasing temperature and shows similar temperature dependence.

Finally, we present the data obtained from the bilayer graphene/NbSe$_2$ and the eight-layer graphene/NbSe$_2$ devices in Figs. 6, The device photographs of each devices are shown in Figs. 6(a) and 6(e), respectively. The $I$-$V$ curves shown in Figs. 6(b) and 6(f) measured at different $V_{BG}$ values demonstrate gate modulation of $I_c$ for bilayer graphene/NbSe$_2$ and eight-layer graphene/NbSe$_2$ device, respectively. The $V_{BG}$ dependences of $I_c$ for each device are summarized in Figs. 6(c) and 6(g), respectively. These change of $I_c$ are showing good coincidence with the $V_{BG}$ dependence of the resistance of bilayer graphene or eight-layer graphene electrode shown in Figs. 6(d) and 6(h), respectively. The smallest critical current density demonstrated in bilayer-graphene/NbSe$_2$ and the eight-layer graphene/NbSe$_2$ devices are calculated to be $1.7 \times 10^4$ and $3.3 \times 10^4$ A/cm$^2$, respectively. These results demonstrate that not only the mono-layer graphene electrode, but also bilayer or few-layer graphene electrode exhibits significant heat transfer effect across the vdW interface. We think this is a consequence of the small electron-phonon coupling of both monolayer graphene and few-layer graphene [26]; thus their electron temperature can be easily increase under the application of current [27] and influence the superconductivity of adjacent $NbSe_2$ layer. In total, we have measured three monolayer graphene/$NbSe_2$ devices, one bilayer graphene/$NbSe_2$ device, and one eight-layer graphene/$NbSe_2$ device. In all the devices, we observed significant reduction of $I_c$ in $NbSe_2$ flake compared to its bulk value as well as the gate modulations of $I_c$ when current is applied from graphene electrode; therefore, these results demonstrate the robustness of the heat transfer effect at graphene/$NbSe_2$ vdw interface.

## 4. Conclusion

We performed systematic comparisons of the superconducting critical current of an $NbSe_2$ flake when current is applied with a Au/Ti contact and with a graphene vdW contact. We found that the superconducting critical current is significantly reduced when the current is applied with the graphene vdW contact compared with that applied with the Au/Ti contact. Moreover, we demonstrated that the critical current of $NbSe_2$ is significantly altered by the gate voltage when the current is applied with the graphene vdW contact. These results are attributed to the significant heat transfer from the graphene electrode under the application of current. When using graphene as an electrode, it is necessary to consider the potential change of the superconducting property of $NbSe_2$ caused by the increase in the electron temperature of the graphene.

**Acknowledgments**

This work was supported by CREST Grant Number JPMJCR15F3, Japan Science and Technology Agency (JST) and JSPS KAKENHI Grant Numbers JP25107003, JP26248061, JP15H01010, JP16H00982, JP18K14083.

**Appendix A: Estimation of contact resistances at Ti/graphene and NbSe₂/graphene junctions**

To estimate the contact resistance contribution of the Au/Ti/graphene and the NbSe₂/graphene junctions, the following procedure has been used. The two-terminal resistance of graphene $R_{2T}$ vs. $V_{BG}$ are measured under the magnetic field of 8.5 T applied perpendicular to the plane and data is shown in Fig. 7(a). (Measurement configuration is same as shown in Fig. 3(d) of the main text). Well-defined quantum Hall plateaus are visible at the $V_{BG}$ values of -4, 7, 17, and 27 V. These corresponds to a filling factor $\nu = 2$, 6, and 10, respectively. Other plateaus were not clearly visible and thus did not used for the analysis. Within the quantum Hall plateau, $R_{2T}$ can be desisribed as $R_{2T}=(R_K/\nu)+R_c$, where $R_K= 25812.807\ \Omega$ denotes the von Klitzing constant, $\nu$ the filling factor, $R_c$ the total resistances contribution of graphene/NbSe₂ contact, Au/Ti/graphene contact, and lead resistance of Au/Ti. The deviation from $R_K/\nu$ at each of quantum Hall plateau gives us the estimation of $R_c$ and it is plotted in Fig. 7(b). The determined $R_c$ value does not significantly change with the $V_{BG}$.

**Appendix B: *I-V* curves of NbSe₂ flake obtained from different voltage probe configurations**

*I-V* curves measured on NbSe₂ flakes are presented in Fig. 8. The contact electrode numbers are shown in Fig. 1(a). In Fig. 8(a), *I-V* curves are displayed for the current applied with graphene electrode (dc current is applied between contacts 3 and 1 in Fig. 1(a)). Different pairs of contacts are used for measuring voltage. In Fig. 8(b), *I-V* curves are displayed for current applied with the metal electrode (dc current is applied between contacts 3 and 6 in Fig. 1(a)). Similar to Fig. 1(b), voltage measured with different pairs of contacts are displayed. Both Figs. 8(a) and 8(b) show that the critical current $I_c$ for the breakdown of superconductivity does not depend on the measurement voltage configuration.

**Figure captions**

**Figure 1** (a) Optical micrograph of the fabricated device. (b,c) Schematic illustration of the graphene/NbSe$_2$ device structure with different current application geometries. Arrows indicate the flow of current. (b) Current application with graphene electrode. (c) Current application with Au/Ti metal electrode. (d) Diagram of heat flow in the device. $p$ denotes the injected power density to the electrons of the graphene due to Joule heating, $G_{\mathrm{e}}^{\mathrm{vdW}}$ the heat conduction through electron at graphene/NbSe$_2$ vdW interface, $G_{\mathrm{ep}}^{\mathrm{Gr}}$ the heat conduction from electrons to phonons in graphene, $G_{\mathrm{p}}^{\mathrm{vdW}}$ the heat conduction through phonons at graphene/NbSe$_2$ vdW interface, $G_{\mathrm{ep}}^{\mathrm{NS}}$ the heat conduction between electrons to phonons in NbSe$_2$, and the heat conduction from graphene or NbSe$_2$ to SiO$_2$/Si substrate heat bath via phonons. (e) Schematic illustration of graphene/NbSe$_2$ vdW interface.

**Figure 2** (a) Current–voltage ($I$–$V$) curves for different measurement configurations; current application with the graphene electrode or application with the metal electrode at $V_{\mathrm{G}} = -50$ V. Data obtained from temperatures $T = 2.0$ and 7.8 K are presented. Measurements are performed under zero magnetic fields. (b) Critical current $I_{\mathrm{c}}$ values for the current application with the graphene electrode at 2 K with different voltage probe configuration. (c) Critical current $I_{\mathrm{c}}$ values for the current application with the metal electrode at 2 K and $V_{\mathrm{BG}} = -50$ V with different voltage probe configurations.

**Figure 3** (a) $I$–$V$ curves at 2 K at various back-gate voltages $V_{\mathrm{BG}}$ measured with current application with the graphene (Gr.) electrode. The curves are offset for clarity and dashed lines indicate offset values. (b) $I$–$V$ curves at 2 K at various $V_{\mathrm{BG}}$ measured with current application with the metal electrode. The curves are offset for clarity and dashed lines indicate offset values. (c) $V_{\mathrm{BG}}$ dependence of critical current $I_{\mathrm{c}}$ for current application with the graphene contact. (d) $V_{\mathrm{BG}}$ dependence of the two-terminal resistance of the graphene measured at 2.0 K. Inset: illustration of the measurement configuration. (e) $V_{\mathrm{BG}}$ dependence of critical current $I_{\mathrm{c}}$ for current application with the metal contact.

**Figure 4** (a) The measurement geometry of the voltage probes. (b) $I$–$V$ curves at 2 K at various back-gate voltages $V_{\mathrm{BG}}$ measured with current application with the graphene electrode. The curves are offset for clarity and dashed lines indicate offset values. (c) Current $I$ dependence of the power $P$ injected to the device measured at different $V_{\mathrm{BG}}$. The curves are offset for clarity and dashed lines indicate offset values. (d) Plot of $I$ values for constant injected powers of $P$ = 25, 50, and 75 μW. In comparison, $V_{BG}$ dependence of $I_c$ is plotted with circles.

**Figure 5** (a-c) $I$–$V$ curves obtained at different temperatures of 2.0 K, 4.8 K, 5.8 K, and 7.8 K for current application with the graphene electrode at different $V_{BG}$ values of (a) −50 V, (b) −15 V, and (c) 0 V. (d-f) $P$–$V$ curves obtained at different temperatures of 2.0 K, 4.8 K, 5.8 K, and 7.8 K for current application with the graphene electrode at different $V_{BG}$ values of (d) −50 V, (e) −15 V, and (f) 0 V. Critical power $P_c$ values for positive $P$ are indicated by arrows. (g) Critical power $P_c$ with respect to the temperature $T$ obtained for different values of $V_{BG}$.

**Figure 6** (a-d) Data obtained from bilayer graphene(BLG)/51-nm-thick NbSe$_2$ device. (a) Optical micrograph. (b) $I$−$V$ curves at 2 K at various back-gate voltages $V_{BG}$ measured with current application with the bilayer graphene electrode. (c) $V_{BG}$ dependence of critical current $I_c$ for current application with the bilayer graphene contact at 2.0 K. (d) $V_{BG}$ dependence of the resistance of the bilayer graphene measured at 2.0 K. (e-h) Data obtained from eight-layer graphene(8LG)/40-nm-thick NbSe$_2$ device. (e) Optical micrograph. (f) $I$−$V$ curves at 2 K at various back-gate voltages $V_{BG}$ measured with current application with the eight-layer graphene electrode. (g) $V_{BG}$ dependence of critical current $I_c$ for current application with the eight-layer graphene contact at 2.0 K. (h) $V_{BG}$ dependence of the resistance of the eight-layer graphene measured at 2.0 K.

**Figure 7** (a) Two-terminal differential resistance $dV/dI$ of a graphene as a function of $V_{BG}$ measured with the geometry shown in Fig. 3(d). Measurement temperature was 2 K and magnetic field of 8.5 T was applied perpendicular to the sample plane. (b,c) Closeup of the Fig. 7(a). (d) Total contact resistance contribution $R_c$ of the Au/Ti/graphene and the NbSe$_2$/graphene junctions determined for different $V_{BG}$ values.

**Figure 8** (a) Current–voltage ($I$–$V$) curves for different voltage measurement configurations for current application with the graphene electrode at $V_G$ = −50 V at 2.0 K. (b) Current–voltage ($I$–$V$) curves for different voltage measurement configurations for current application with the metal electrode at $V_{BG}$ = −50 V at 2.0 K. Arrows indicate sweep direction of current $I$.

**Figure 1**

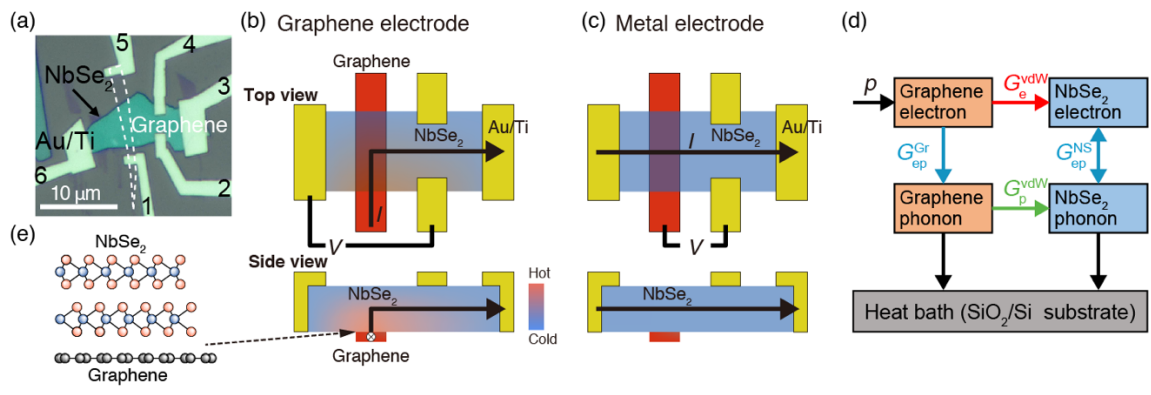

**Figure 2**

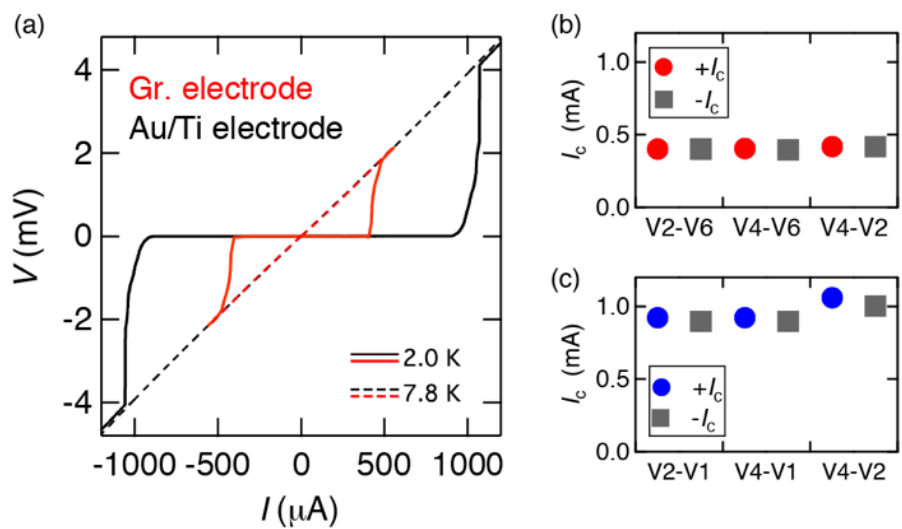

**Figure 3**

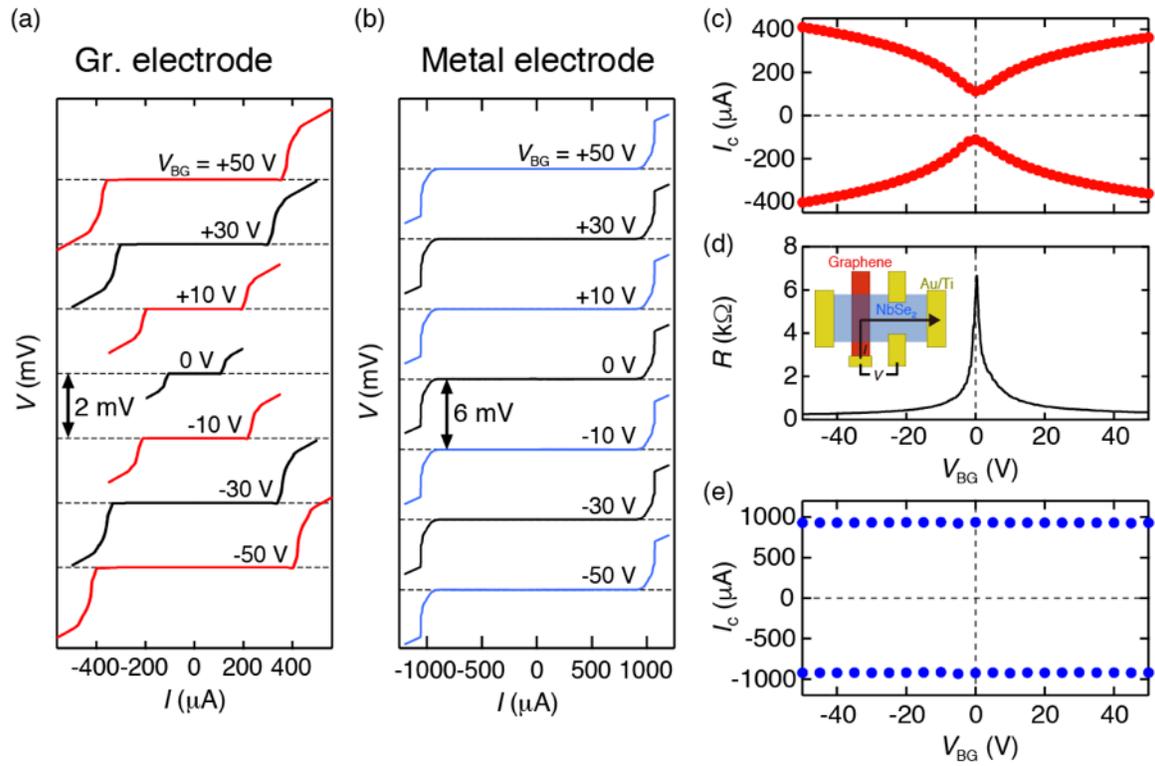

**Figure 4**

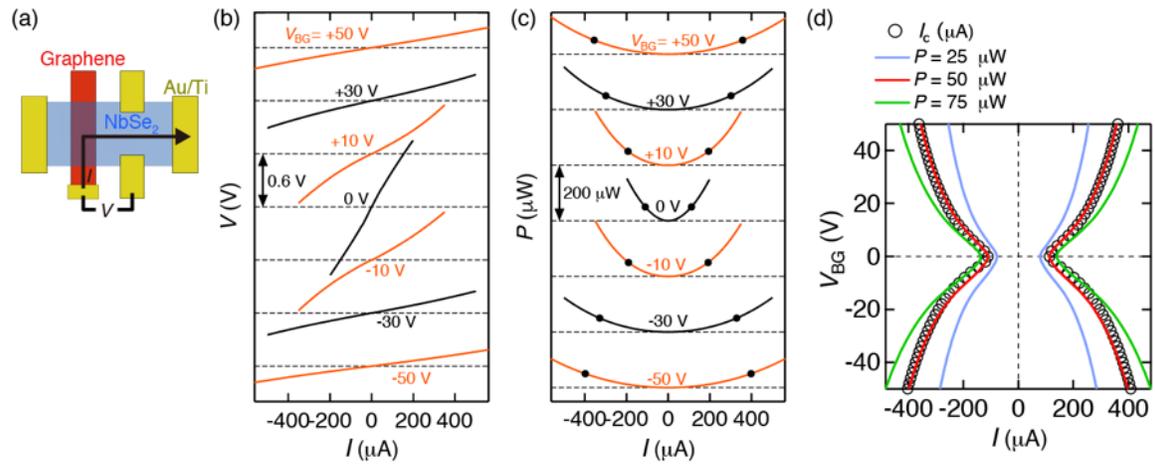

**Figure 5**

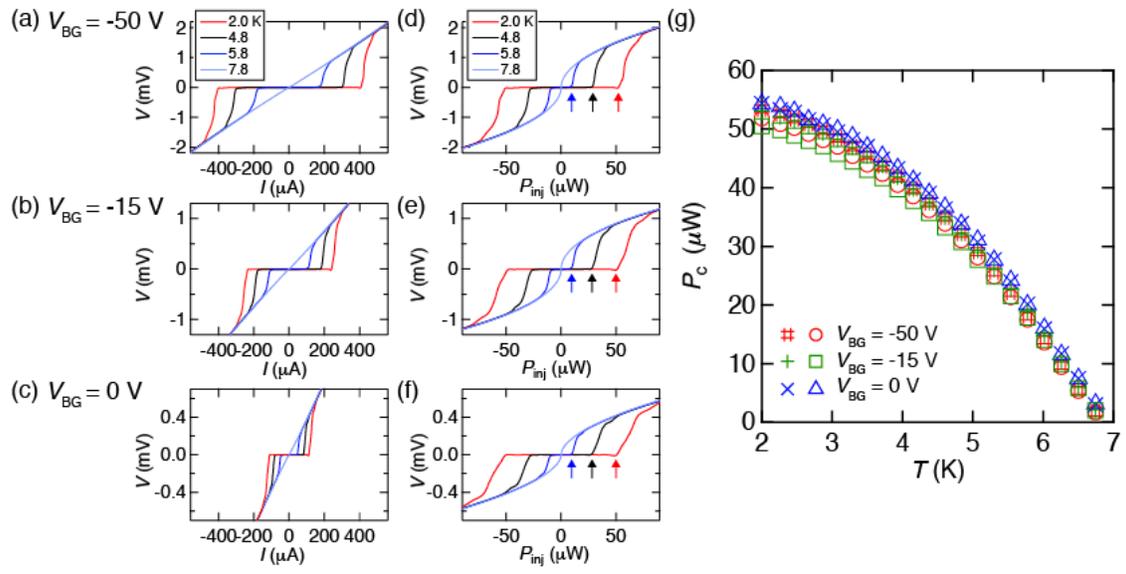

**Figure 6**

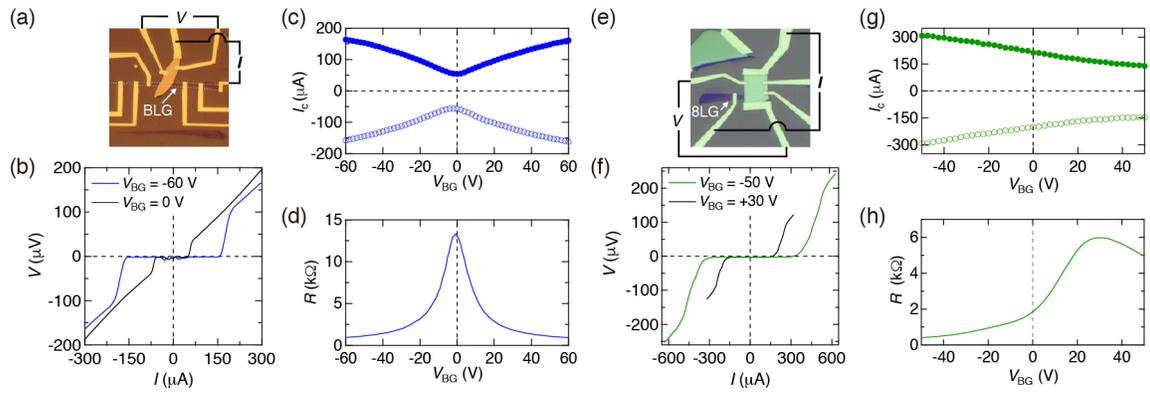

**Figure 7**

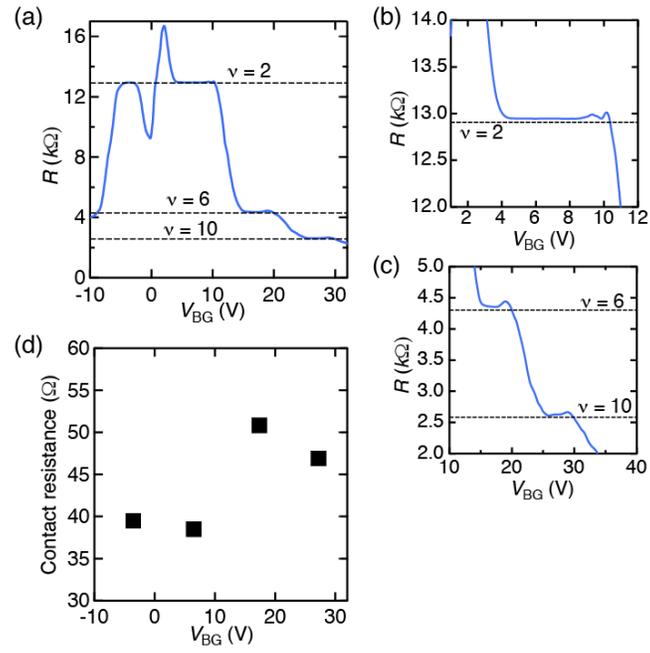

**Figure 8**

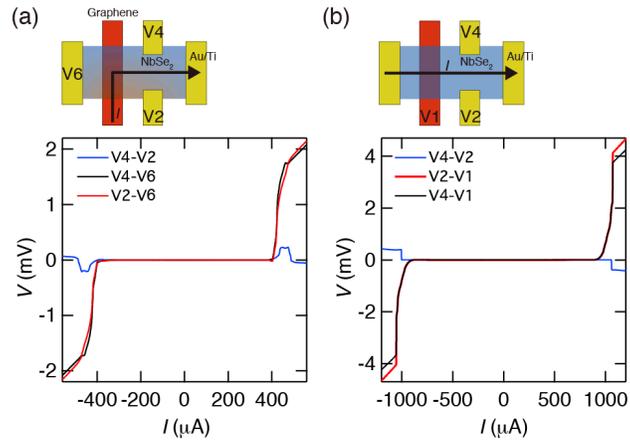